\documentclass[manuscript]{emulateapj}

\usepackage{longtable, lscape}

\begin{document}

\title{Rotational Line Emission from Water in Protoplanetary Disks}

\author{R. Meijerink\altaffilmark{1}, D.R. Poelman\altaffilmark{2}, M.
  Spaans\altaffilmark{3}, A.G.G.M. Tielens\altaffilmark{4,3}, and A.E.
  Glassgold\altaffilmark{1}}

\altaffiltext{1}{Astronomy
  Department, University of California, Berkeley, CA 94720, United
  States}\email{rowin@astro.berkeley.edu}
\altaffiltext{2}{School of Physics \&
  Astronomy, University of St Andrews, North Haugh, St Andrews KY16
  9SS, Scotland}
\altaffiltext{3}{Kapteyn Astronomical
  Institute, P.O. Box 800, 9700 AV Groningen, The Netherlands}
\altaffiltext{4}{NASA Ames Research Center, MS245-3,
  Moffett Field, CA 94035, USA}

\begin{abstract}
  Circumstellar disks provide the material reservoir for the growth of
  young stars and for planet formation. We combine a high-level
  radiative transfer program with a thermal-chemical model of a
  typical T Tauri star disk to investigate the diagnostic potential of
  the far-infrared lines of water for probing disk structure. We
  discuss the observability of pure rotational H$_2$O lines with the
  {\it Herschel Space Observatory}, specifically the residual gas
  where water is mainly frozen out. We find that measuring both the
  line profile of the ground $1_{10}-1_{01}$ ortho-H$_2$O transition
  and the ratio of this line to the $3_{12}-3_{03}$ and
  $2_{21}-2_{12}$ line can provide information on the gas phase water
  between 5-100~AU, but not on the snow line which is expected to
  occur at smaller radii.

\end{abstract}


\section{Introduction}

The radius where volatiles sublimate in a protoplanetary disk plays an
important role in planet formation. One volatile of particular
interest is water, and the place where it sublimates or freezes out
onto grains is called the {\it snow line} \citep{Hayashi1981}. The
sublimation temperature for water ranges from $T\sim 110-170$~K,
depending on the water vapor pressure (cf. Fraser et
al.~\citeyear{Fraser2001}; Podolak \& Zucker~\citeyear{Podolak2004}).
At radii larger than the snow line, it is possible to hydrate
planetesimals and protoplanets.  \citet{Hayashi1981} found that the
snow line is now located at about 2.7~AU from the Sun in our own Solar
System. However, the position can be different for a T Tauri star. Its
location depends on the evolutionary phase of the disk, and it can
vary from 0.7 to 10~AU \citep{Sasselov2000, Lecar2006, Garaud2007}.

Gas phase water is expected to be abundant between the water-ice
sublimation temperature $T\sim 110-170$~K and the dissociation
temperature $T\sim 2500$~K. Its rich spectrum covers almost all
astronomically available wavelength bands.  Using the near-infrared
lines near 2.3$\mu$m, \citet{Carr2004} and \citet{Thi2005} found warm
H$_2$O in protoplanetary disks at temperatures $T>1200$~K and radii
$R<0.4$~AU, and \citet{Carr2008} and \citet{Salyk2008} have detected
mid-infrared water lines with {\it Spitzer}. The far-infrared
spectrometers aboard the {\it Herschel} space telescope will offer the
opportunity to measure the pure rotational transitions of this
important molecule, which are characteristic of gas temperatures of
$T\gtrsim 60$~K.  Because gas phase H$_2$O requires warm conditions
($T\gtrsim 200$~K) it is expected to be abundant at small radii and
thus perhaps not observable with the {\it Herschel} spectrometers.
However, temperatures appropriate to both the occurrence and detection
of H$_2$O are present over a wide range of radii because stellar
radiation heats the disk atmosphere.  Moreover, as we show in this
letter, several lines of H$_2$O can be detected by {\it Herschel} at
very low abundances close to the mid-plane where the bulk of the water
is frozen. The detection of these lines will complement the previous
measurements of the near-infrared and mid-infrared water lines.

\section{Models}

The calculation is done as follows: (1) We use an X-ray irradiated disk 
code to calculate the temperature structure and molecular abundances 
for a generic T Tauri disk. (2) The results are input into a multi-zone 
radiative transfer code that includes the calculation of the excitation 
of the H$_2$O lines. (3) A ray-tracing code is used to obtain line fluxes 
and line shapes. These three steps are discussed in more detail below.

\subsection{The thermal and chemical structure of the disk}

\begin{figure}[!htp]
\centerline{\includegraphics[height=45mm,clip=]{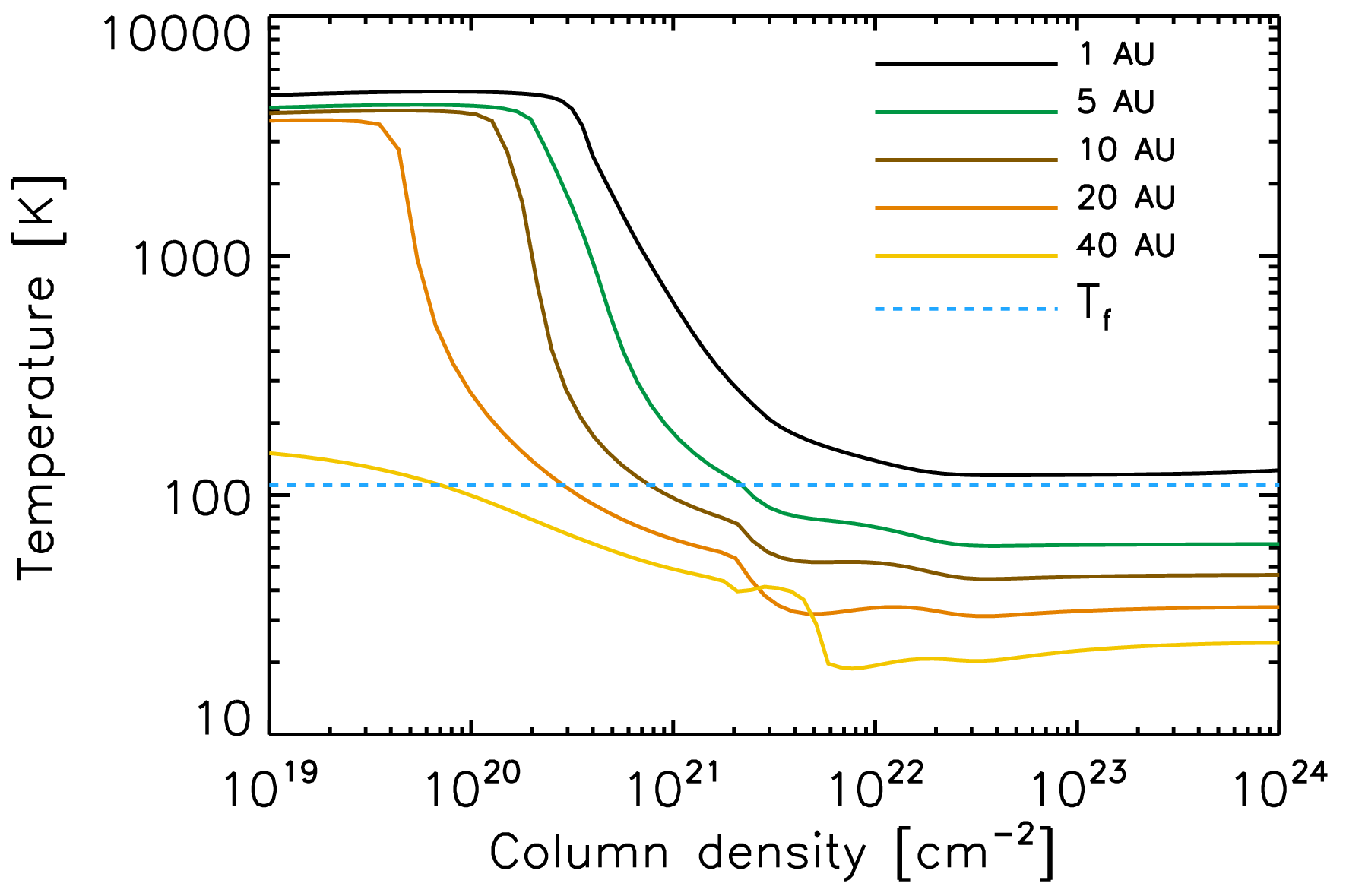}}
\centerline{\includegraphics[height=45mm,clip=]{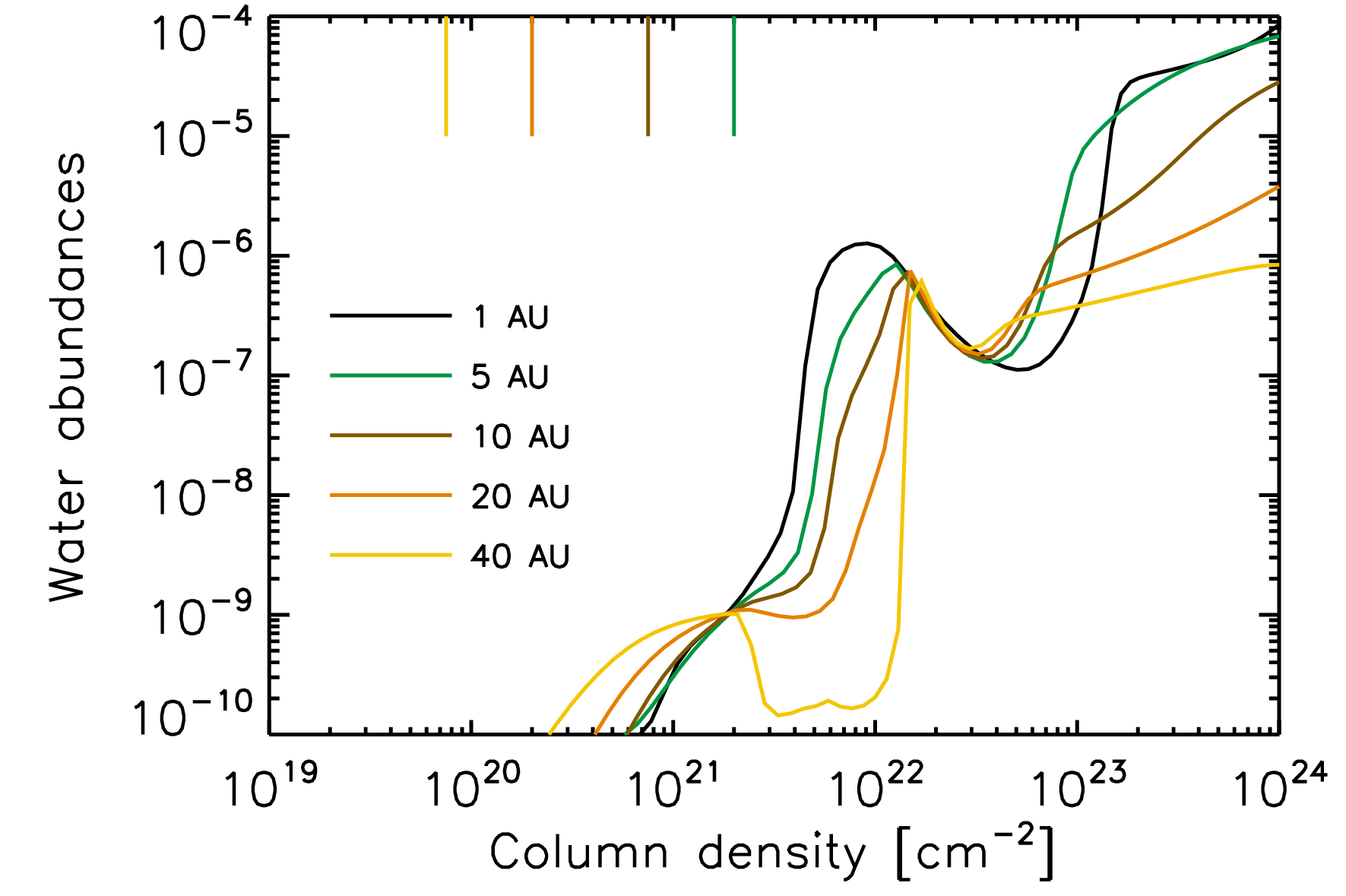}}
\caption{Temperature (top), and fractional water abundance (bottom)
  vs. perpendicular column density. The dotted blue line (top) shows
  the freeze-out temperature $T_f=110$~K. The colored tick marks at
  the top of the bottom panel indicate the smallest perpendicular
  column density for each radius where water freezes out. No
  freeze-out occurs within $R=1$~AU. Note the non-monotone variation
  of the water abundance due to gradients in the ionization rate,
  density, and temperature.}
\label{temp_and_abun}
\end{figure}

The thermal-chemical structure of the disk is calculated with the code
described by \citet{Glassgold2004} with minor corrections and updates
\citep{Meijerink2008}. The disk is illuminated by stellar X-rays with
a thermal spectrum with temperature $T_X=1$~keV and luminosity
$L_X=2\times10^{30}$~erg~s$^{-1}$. In regions where the X-rays are
strongly attenuated, the disk is ionized by the decay products of
$^{26}$Al at a rate $\zeta_{26}=4\times10^{-19}$~s$^{-1}$. The density
structure is given by the generic T Tauri disk model (Paola d'Alessio,
private communication) with accretion rate $\dot{M} = 10^{-8}\,{\rm
  M}_{\sun}$\,yr$^{-1}$ and stellar parameters $M_*=0.5$~M$_\odot$,
$R_*=2{\rm R}_\odot$, and $T_*=4000$~K. The disk is flared, and the
density varies continuously from $R = 0.028$ to $>$500 AU. The model
density does not include modifications such as holes, gaps, and rims
suggested, for example, by the spectral energy distributions measured
with {\em Spitzer} (e.g., Dullemond et
al.~\citeyear{Dullemond2007}).The adopted model has a power law grain
size distribution with index $p=-3.5$ and minimum and maximum grain
sizes of $a_{min}=0.005$ and $a_{max}=1000$~$\mu$m, respectively,
resulting in a geometric mean grain size, $a_g=2.24$~$\mu$m. The grain
size $a_g$ determines the thermal coupling between dust and gas.

The \citet{dAlessio1999,dAlessio2001} dust temperature involves a
balance between heating by viscous dissipation, FUV/optical/IR
radiation, cosmic rays and cooling by IR radiation. Our gas
temperature is determined by heating by penetrating X-rays, gas-grain
collisions and cooling by atomic and molecular lines.  The gas thermal
model relies solely on X-ray irradiation in contrast to UV irradiation
models by \citet{Jonkheid2004,Jonkheid2007}, \citet{Kamp2004}, and
\citet{Nomura2007} (who also include X-rays), who include UV photons.
The UV irradiation, however, play no role in the regions where the
submillimeter H$_2$O emission is produced. The numerical approach is
"1+1D", i.e., the vertical thermal and chemical structure is
calculated independently at 45 radial points whose spacing increases
with radius.  However, the X-ray attenuation is calculated along the
line-of-sight to the star.  The photons can only escape in the
vertical direction, a reasonable approximation, since the opacities in
the radial direction are so much larger.

The chemistry is steady state and entirely gas phase, and it includes
only carbon and oxygen bearing species. It does not treat freeze-out
of H$_2$O in detail, but we expect that freeze-out cannot be 100\%
efficient.  Incomplete freeze-out has been observed by ISO
\citep{Boonman2003}, and \citet{Davis2007} has modeled freeze-out in
irradiated accretion disks. In order to treat the freeze-out of water,
we assume a fixed gas-phase water abundance, $x_f({\rm H_2O})=10^{-8}$
or $x_f({\rm H_2O})=10^{-10}$, in regions where most of the water is
expected to freeze-out on grains, i.e., where the gas temperature
satisfies the condition $T<T_f$ with $T_f=110$~K, the adopted
freeze-out temperature. Where the model gas phase abundance is lower,
we adopt the calculated abundance. A more realistic treatment of water
freeze-out would involve treating many poorly understood physical
processes beyond the scope of this short report.  These might include
the growth and settling of the dust grains, molecular synthesis and
transformation on grain surfaces, adsorption and desorption, and
mixing. We are developing an improved thermal-chemical disk model that
will address many of these issues.  The advantage of the present model
is that it allows us to isolate the chemical effects of freeze-out and
the effects of grain growth and settling in a simple albeit
preliminary way. 

\subsection{Water radiative transfer}

The critical densities of the lines of interest are typically large,
in the range $n_{\rm crit}\sim10^8-10^{10}$~cm$^{-3}$. The fractional
abundance of gaseous water becomes significant at perpendicular column
densities $N_{\rm H}> 2-5\times 10^{21}$~cm$^{-2}$
(Fig.\ref{temp_and_abun}), and the excitation can be sub-thermal,
especially at large radii ($R>10-20$~AU). Thus assuming local
thermodynamic equilibrium is not always a good approximation, and a
full radiation transfer calculation is required.

We calculate the level populations of H$_2$O using the multi-zone
escape probability method $\beta$3D described by
\citet{Poelman2005,Poelman2006}. We assume Keplerian rotation for a
stellar mass $M_*=0.5$~M$_\odot$ and turbulent velocities $\delta
v$=0.5, and 2.0 km~s$^{-1}$ (where $\delta v$ is the $1/e$ half width of line
profile). In the multi-zone formalism, the medium is divided into a
large number of grid cells, each with calculated values of the
volumetric hydrogen density, the temperatures of gas and dust, and the
H$_2$O abundance. The connection of every grid cell with all the
others enables us to solve locally the equations of statistical
equilibrium taking into account global information for the entire
disk. Radiative and collisional excitation and de-excitation by
electrons and atomic and molecular hydrogen, and dust pumping are
included, using data from \citet{Faure2008} and the Leiden Atomic and
Molecular Database (LAMDA, Sch\"oier et al.~\citeyear{Schoeier2005}).
We adopt the dust opacities from \citet{dAlessio2001}. Although the
radiative transfer code can deal with three dimensions, we only
consider 1D slabs as in the thermal-chemical calculation.

\subsection{Ray tracing}

We compute line profiles for inclinations ranging from $0^\circ$ to 
$90^\circ$ with the sky brightness distribution program SKY, which is 
part of the RATRAN code \citep{Hogerheijde2000}. We use the 2-dimensional 
cylindrically-symmetric version of SKY, and the resulting level populations, 
densities and molecular abundances are combined and rebinned into a single
grid at each radius. The dynamical ranges of the physical and chemical 
properties are large in the vertical direction, and we use 400 cells in this
direction and decrease the cell size toward the disk midplane. Fewer
grid cells are needed in the radial direction since the changes are more 
gradual, although smaller spacings are also used at small radii.

\section{Results \& Discussion}

Fig. \ref{temp_and_abun} shows the gas temperature and H$_2$O
abundance for a range of radii from $R=1-40$\,AU plotted versus
increasing perpendicular column density $N_{\rm H}$. As explained in
more detail in \citet{Glassgold2004} and \citet{Meijerink2008}, the
stellar X-rays heat the gas at the top of the atmosphere as high as
$T\sim3000-4000$~K for $R<20$~AU.  Attenuation of the X-rays for
$N_{\rm H} >10^{21}$~cm$^{-2}$ produces a drop in temperature and
induces transitions from atoms (or ions) to molecules. For example,
the H to H$_2$ transition occurs for $N_{\rm
  H}=10^{21}-10^{22}$~cm$^{-2}$. The water abundance before the H to
H$_2$ transition is small due to the efficient destruction by X-ray
ionized species such as H$^+$ and He$^+$. Congruent with the H to
H$_2$ transition, the water abundance increases steeply.

According to Fig.~\ref{temp_and_abun}, freeze-out occurs for $R >
1$\,AU and beyond $N_{\rm H}\sim 10^{19}-10^{22}$~cm$^{-2}$ depending
on the radius (as indicated by the colored tick marks for each
radius). The volumetric densities here are quite high, $n_{\rm
  H}=10^{8}-10^{12}$~cm$^{-3}$ and, despite the reduced dust surface
area in the thermal-chemical model, the gas is thermally coupled to
the dust. At even higher column densities than are shown in Fig.
\ref{temp_and_abun} (closer to the mid-plane), the temperature
increases slightly due to viscous accretion heating. Temperatures
greater than $T\sim200$~K are needed for efficient gas phase synthesis
of water by neutral radical reactions, and they are only attained at
small radii $R < 0.5$~AU, where $x_{\rm H_2O}>10^{-4}$ for N$_{\rm H}
> 10^{22}-10^{23}$~cm$^{-2}$.

We have calculated line profiles and fluxes for the low-excitation
H$_2$O transitions that are in the spectral range of {\it Herschel}
{\it Photodetector Array Camera and Spectrometer} (PACS) and {\it
  Heterodyne Instrument on the Far-Infrared} (HIFI). The results
indicate that a number lines in the frequency bands of HIFI are
detectable. The frequency bands of PACS are less favorable, and the
following discussion is based on HIFI. The integrated fluxes are
summarized in Table \ref{rotational_fluxes} for sources at an assumed
distance of 140~pc. During the first {\it Herschel} observation cycle,
only the ortho $1_{10}-1_{01}$ and para $1_{11}-0_{00}$ ground level
transitions will be observed in the {\it Herschel} key program "Water
in Star-forming Regions With {\it Herschel}"
(http://www.strw.leidenuniv.nl/WISH/), and we discuss these lines
first.

\begin{table}[htp]
\caption[]{ Integrated intensities $I = \int T_{MB}dv$[mK km s$^{-1}$]$^a$}
\centering
\begin{tabular}{lccccccc}
  \tableline
  Trans.  & Freq. & \multicolumn{2}{c}{No freeze-out} &  \multicolumn{2}{c}{$x_f=10^{-8}$} &  \multicolumn{2}{c}{$x_f=10^{-10}$} \\
          & [Hz]  & \multicolumn{2}{c}{} &  \multicolumn{2}{c}{} & \multicolumn{2}{c}{}  \\
  \tableline
              &           & \multicolumn{6}{c}{Turbulent velocity $\delta v$ [km~s$^{-1}$]$^b$} \\
              &           &  0.5 & 2.0 & 0.5 & 2.0 & 0.5 & 2.0 \\
  \tableline
  \multicolumn{8}{c}{ortho-H$_2$O lines} \\
  $1_{10}$-$1_{01}$ & 556.936 & 13.7 &  66.2 & 16.3 &  59.3 & 13.6 & 41.8 \\
  $3_{12}$-$3_{03}$ & 1097.37 & 32.7 & 120.9 & 19.1 &  50.6 &  3.9 &  7.0 \\
  $3_{12}$-$2_{21}$ & 1153.13 & 33.5 & 103.6 & 12.5 &  29.0 &  1.4 &  1.9 \\
  $3_{21}$-$3_{12}$ & 1162.91 & 27.6 &  94.6 & 12.9 &  32.5 &  2.4 &  4.3 \\
  $2_{21}$-$2_{12}$ & 1661.01 & 34.3 & 143.7 & 29.8 &  97.2 & 10.9 & 20.5 \\
  $2_{12}$-$1_{01}$ & 1669.90 & 21.1 & 114.2 & 26.6 &  97.0 & 25.5 & 85.5 \\
  $3_{03}$-$2_{12}$ & 1716.77 & 38.9 & 150.2 & 32.3 & 110.1 & 15.5 & 33.2 \\
  $3_{30}$-$3_{21}$ & 2196.35 & 25.3 &  98.1 & 11.5 &  27.9 &  2.1 &  3.5 \\
  $4_{14}$-$3_{03}$ & 2640.47 & 28.6 & 125.3 & 25.3 &  73.0 &  9.2 & 20.0 \\
  $2_{12}$-$1_{10}$ & 2773.97 & 16.1 &  96.2 & 21.4 &  69.1 & 17.0 & 51.0 \\
  \tableline
  \multicolumn{8}{c}{para-H$_2$O lines} \\
  $2_{11}$-$2_{02}$ & 752.029 & 27.6 & 108.7 & 21.3 &  62.5 &  4.9 &  7.9 \\
  $2_{02}$-$1_{11}$ & 987.924 & 39.9 & 138.9 & 29.8 & 102.4 & 10.6 & 17.3 \\
  $1_{11}$-$0_{00}$ & 1113.34 & 21.4 & 105.5 & 24.9 &  94.5 & 23.1 & 70.0 \\
  $2_{20}$-$2_{11}$ & 1228.80 & 32.7 & 124.5 & 20.7 &  56.9 &  3.9 &  6.0 \\
  \tableline
\end{tabular}
\tablenotetext{a}{Adopting distance $d=140$~pc}
\tablenotetext{b}{$\delta v$ is the $1/e$ half width of line profile.}  
\label{rotational_fluxes}
\end{table}

\subsection{The $1_{10}-1_{01}$ and $1_{11}-0_{00}$ H$_2$O lines}

\begin{figure}[htp]
\centerline{\includegraphics[height=60mm,clip=]{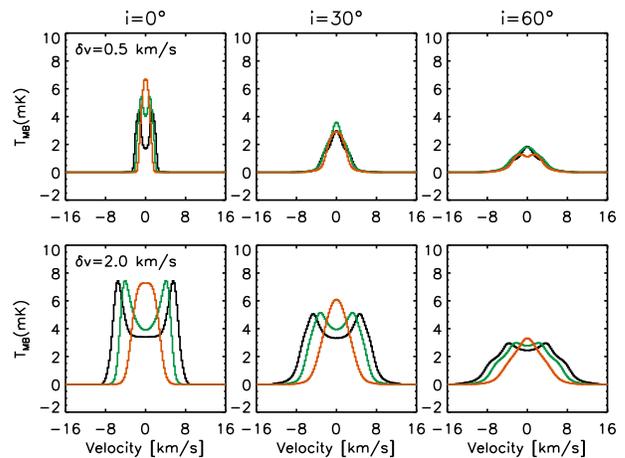}}
\caption{ The $1_{10}-1_{01}$ ortho-H$_2$O lines with the d'Alessio
  opacities for inclinations of 0$^\circ$ (left), 30$^\circ$ (middle),
  and 60$^\circ$ (right); no freeze-out (black), with freeze-out below
  110~K and abundance $x_f({\rm H_2O})=10^{-8}$ (green) and $x_f({\rm
    H_2O})=10^{-10}$ (red) in the freeze-out zone; turbulent velocity
  $\delta v=0.5$(top), and 2.0~km~s$^{-1}$ (bottom).}
\label{over_lay_110_101}
\end{figure}

The line profiles of the 556.936 GHz $1_{10}-1_{01}$ ortho and the
1113.34 GHz $1_{11}-0_{00}$ para transitions are convolved with the
38\arcsec\ and 19\arcsec {\it Herschel} beam sizes at these
frequencies.  Fig. \ref{over_lay_110_101} shows how the profile of the
$1_{10}-1_{01}$ ortho transition is affected by changes in inclination
angle $i$, turbulent velocity $\delta v$, and the effects of
incomplete freeze-out. Variations in $i$ are shown horizontally across
the figure, and changes in $\delta v$ vertically; the three lines
represent no freeze-out and freeze-out that leaves residual water
abundances of $10^{-8}$ and $10^{-10}$. The integrated intensities for
the models shown vary between 14 and 66 mK~km~s$^{-1}$, and are mainly
sensitive to the line width. The $1_{10}-1_{01}$ ortho- and
$1_{11}-0_{00}$ para-H$_2$O lines have similar integrated intensities,
with line ratios $1_{10}-1_{01}/1_{11}-0_{00} \sim 0.6$. The small
variations in this ratio, caused by high opacities in the lines, makes
these lines unfavorable as diagnostics.

Due to the rotation of the disk, the lines are broadened with
increasing inclination, but at increasingly smaller integrated
intensities because of the reduction of the apparent surface area of
the disk. For a completely edge on disk, absorption by the cold outer
regions ($R\gtrsim 100$~AU) is important, but this effect has not been
included in the present calculations. At an inclination of 60$^\circ$,
the rotational broadening extends to 5~km~s$^{-1}$ (best seen at
$\delta v = 0.5$~km~s$^{-1}$), which corresponds to a rotational
velocity of $v_r=10$~km~s$^{-1}$ or a radius of $R=10$~AU. The
dominant contributions to the $1_{10}-1_{01}$ line flux originate
therefore from $R\gtrsim 10$ to 100~AU. Thus this line provides no
information on the snow line, which is located close to the star, near
$R\sim 1$~AU in this model (see Fig. \ref{temp_and_abun}).

At an inclination $i=0^\circ$ (face-on disk), the line width is
determined by the turbulent velocity $\delta v$ and the line opacity.
The center of the line is strongly self-absorbed for no freeze-out,
since the emission produced at $N_{\rm H}\sim
10^{22}-10^{23}$~cm$^{-2}$ is absorbed by the upper layers. These
regions are sub-thermally excited because the densities are low
($n_{\rm H}<10^8$~cm$^{-3}$). Self-absorption occurs, but to a lesser
extent, when freeze-out is included and the residual water vapor
abundance is $x_f$(H$_2$O)=$10^{-8}$. Even for
$x_f$(H$_2$O)=$10^{-10}$, the line is still optically thick. For
finite inclination angles, $i=30^\circ$ and $60^\circ$, flux loss at
line center can also occur for the freeze-out model with a water vapor
abundance $x_f({\rm H_2O})=10^{-10}$. This is most clearly seen for
the case $\delta v= 0.5$~km~s$^{-1}$. This is not a signature of the
snow line, since the flux loss occurs for $i=60^\circ$, for example,
for velocities $v<2$~km~s$^{-1}$, corresponding to radii $R>50$~AU.

\begin{figure}[htp]
\centerline{\includegraphics[height=30mm,clip=]{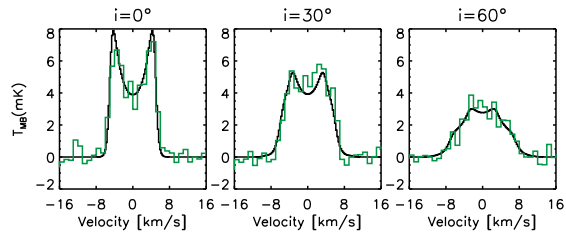}}
\caption{The $1_{10}-1_{01}$ ortho-H$_2$O line profile with a
  turbulent velocity of 2.0~km~s$^{-1}$ for inclinations (from left to
  right) $i=0^\circ$, 30$^\circ$, 60$^\circ$, with freeze-out and
  $x_f({\rm H_2O})=10^{-8}$. The model results are in black. For the
  curves in green, a 1$\sigma$ noise level of 0.5~mK has been added.
  The velocity resolution is assumed to be 1~km~s$^{-1}$.  }
\label{noise_added_model}
\end{figure}

In Fig.  \ref{noise_added_model}, we test how well the shape of the
$1_{10}-1_{01}$ line can be measured by {\it Herschel}. The model is
shown in black, and a 1$\sigma$ noise level of 0.5~mK has been added
in green. The sensitivity of the {\it Herschel} detectors at
1113.34~GHz is about four times less than at 556.936~GHz. According to
the time estimator for the {\it Herschel} Space Observatory
(http://www.esac.esa.int), a 1$\sigma$ noise level of 0.5 and 2.0~mK
can be obtained at the frequencies of the $1_{10}-1_{01}$ ortho and
$1_{11}-0_{00}$ para transition, respectively, in about 4 hours of
observing time, when frequency switching is used and a spectral
resolution of 1~km~s$^{-1}$ is requested. Thus one should be able to
determine the width, and to some degree the shape of the
$1_{10}-1_{01}$ ortho-H$_2$O line. However, the effect of
self-absorption due to high opacities will be hard to distinguish for
low turbulent velocities ($\delta v \leq 1.0$~km~s$^{-1}$) at this
resolution. The para line is not a good candidate as the ortho
transition for obtaining a line shape at the same noise level because
of the decrease in sensitivity.

\subsection{Line ratios}

The line profiles and integrated intensities of the $1_{10}-1_{01}$
and its ratio with the $1_{11}-0_{00}$ transition are affected by
inclination angle, turbulent broadening, and the residual water vapor
abundance after freeze-out. The dust opacity, which is influenced by
grain growth and settling, can also play a role, although our study of
these effects indicates that this is less important than the variables
just mentioned. Thus the models are degenerate in the sense that
different parameter sets give the same line shapes and intensities,
due to the high opacities in these lines. Unfortunately, these are the
only lines to be observed in the WISH program. Observing higher
excitation transitions would help, but the decrease in sensitivity
with frequency means that only line fluxes and not line shapes can be
determined by {\it Herschel} observations. At a velocity resolution of
1~km~s$^{-1}$, it would take {\it Herschel} about 4 hours to get the
1$\sigma$ noise level down to 2~mK for transition frequencies around
$\sim 1100$~GHz.  In a similar observation around $\sim 1700$~GHz, the
noise level would be 8~mK.

\begin{figure}[htp]
\centerline{\includegraphics[height=45mm,clip=]{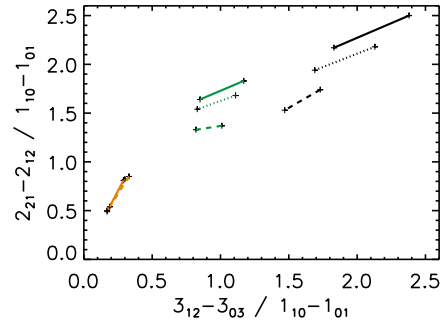}}
\caption{The $3_{12}-3_{03}$/$1_{10}-1_{01}$ versus the
  $2_{21}-2_{12}$/$1_{10}-1_{01}$ ratio for models without freeze-out
  (black), models with gas-phase abundances in the freeze-out zone of
  $x_f({\rm H_2O})=10^{-8}$ (green) and $10^{-10}$ (orange). The
  different lines are for inclinations $i=0^\circ$ (solid), $30^\circ$
  (dotted), and $60^\circ$ (dashed).}
\label{ratios}
\end{figure}

Inspection of Table 1 shows that higher transitions may be observable
with {\it Herschel}. We will now focus on the $3_{12}-3_{03}$
transition at 1097.37~GHz and the $2_{21}-2_{12}$ transition at
1661.01~GHz, although other lines may also have diagnostic value. Fig.
\ref{ratios} plots the ratios of these two lines to the fundamental
$1_{10}-1_{01}$ transition, and shows that the ratios can distinguish
between different levels of water vapor in the freeze-out zone. The
model results for no freeze-out are found in the right upper part of
the figure, where there are three segments for inclination
$i=0^\circ$, $30^\circ$ and $60^\circ$; each segment consists of two
values of the turbulent velocity, $\delta v=0.5$ and
$2.0$~km~s$^{-1}$.  Similar sets for $x_f = 10^{-8}$ (middle part) and
$x_f = 10^{-10}$ (left lower part) are clearly in a different part of
the diagram and do not overlap with each other. The variations due to
inclination and turbulent velocity are smaller than those associated
with freeze-out.

\section{Conclusions}

We have shown that it should be possible to observe some far-infrared
rotational lines of water with the HIFI instrument on {\it Herschel}
that are produced in regions between radii $R\sim 10-100$~AU. The
current scheduled observations of the $1_{10}-1_{01}$ and
$1_{11}-0_{00}$ transitions would provide information on the spatial
distribution of water and on turbulence if the inclination of the disk
is known. The ratios of these lines are quite similar for the
different models ($\sim 0.9-1.0$), which is due to the high opacities
in both lines. The line shapes of the $1_{10}-1_{01}$ transition for
different water vapor residuals in the freeze-out zone are quite
similar at the same inclination angle. When measurements of the
$1_{10}-1_{01}$ line are combined with higher transitions, such as the
$3_{12}-3_{03}$, and $2_{21}-2_{12}$ lines, the abundance of the
residual water vapor in the freeze-out zone can be determined (see
Fig. \ref{ratios}). These {\it Herschel} observations of rotational
lines will complement observations of shorter-wavelength excitation
lines by {\it Spitzer} (10-20$\mu$m, e.g., Carr \& Najita
\citeyear{Carr2008}; Salyk et al. \citeyear{Salyk2008}), SOFIA
(6$\mu$m), and near-infrared ro-vibrational transitions
\citep{Carr2004,Salyk2008}). These lines trace the inner regions of
the disk out to radii $R \sim 1-2$~AU, with the potential of providing
information on the conditions in the immediate neighborhood of the
snow line.

\section{Acknowledgements} 

This work has been supported by NSF Grant AST-0507423 and NASA Grant
NNG06GF88G to UC Berkeley. We would like to express our appreciation
to Dr.~Floris van der Tak for providing the 2D version of SKY, and to
Dr.~Paola d'Alessio for providing the dust opacities. We would also
like to thank the referee for the comments about the dust opacities.



\bibliographystyle{aa} 


\end{document}